\documentclass{article}
\usepackage[tight]{subfigure}
\usepackage[vlined, ruled, linesnumbered]{algorithm2e}
\usepackage{amssymb}
\usepackage{amsmath}
\usepackage{amsthm}
\usepackage{amsfonts}
\usepackage[T1]{fontenc}
\usepackage[cp1250]{inputenc}
\usepackage{graphicx}

\def\pp{2.6488}
\def \eps{\epsilon}
\def \complexity{(9+\epsilon)}
\newcommand{\n}{\mathbb{N}}

\def \algname{Find-Lambda}
\def \complexityy{D\cdot C^{\log y}y^{3\log y}9^y}

\begin{document}
\newtheorem{observation}{Observation}
\newtheorem{theorem}{Theorem}
\newtheorem{definition}{Definition}
\newtheorem{lemma}{Lemma}
\newtheorem{corollary}{Corollary}
\newtheorem{remark}{Remark}
\title{Determining $L(2,1)$-Span in Polynomial Space}

\author{
Konstanty Junosza-Szaniawski,  Pawe{\l} Rz{\k a}{\. z}ewski  \\
\texttt{\{k.szaniawski, p.rzazewski\}@mini.pw.edu.pl} \\ \\
Warsaw University of Technology \\
Faculty of Mathematics and Information Science \\
Pl. Politechniki 1 , 00-661 Warsaw, Poland
}
\date{ }
\maketitle

\begin{abstract}
A $k$-$L(2,1)$-labeling of a graph is a function from its vertex set
into the set $\{0,\dots,k\}$, such that the labels assigned to adjacent
vertices  differ by at least 2, and labels assigned to vertices of
distance 2 are different. It is known that finding the smallest
 $k$ admitting the existence of a $k$-$L(2,1)$-labeling of any given graph is NP-Complete.

In this paper we present an algorithm for this problem,
which works in time $O(\complexity ^n)$ and polynomial memory,
 where $\eps$ is an arbitrarily small positive constant.
This is the first exact algorithm for $L(2,1)$-labeling problem
with time complexity $O(c^n)$ for some constant $c$ and polynomial space complexity.
\end{abstract}

\section{Introduction}
A frequency assignment problem is the problem of assigning channels of
frequency (represented by nonnegative integers) to each radio
transmitter, so that no transmitters interfere with each other.
 Hale \cite{Hab} formulated this problem in terms of so-called $T$-coloring of graphs.

According to \cite{GY}, Roberts was the first who proposed a modification of this problem,
which is called an $L(2,1)$-labeling problem. It asks for such a
labeling with nonnegative integer labels, that no vertices in
distance $2$ in a graph have the same label and labels of adjacent
vertices differ by at least $2$.

A $k$-$L(2,1)$-labeling problem is to determine if there exists an
$L(2,1)$-labeling of a given graph with no label greater than $k$.
By $\lambda(G)$ we denote an $L(2,1)$-span of $G$, which is the smallest
value of $k$ that guarantees the existence of a
$k$-$L(2,1)$-labeling of $G$.

The problem of $L(2,1)$-labeling has been extensively studied (see \cite{Cala,FKsurvey,GK,YehSurvey}
 for some surveys on the problem and its generalizations).
A considerable attention has been given to bounding the value of $\lambda(G)$ by some function of $G$.

Griggs and Yeh \cite{GY} proved that $\lambda(G) \leq \Delta^2 + \Delta$ \footnote{$\Delta$ denotes the largest vertex degree in a graph}
and conjectured, that $\lambda(G) \leq \Delta^2$ for every graph
$G$. There are several results supporting this
conjecture, for example Gon\c{c}alves \cite{goncalves} proved that
$\lambda(G) \leq \Delta^2 + \Delta - 2$ for graphs with $\Delta \geq 3$. Havet {\it et al.}
\cite{Hav} have settled the conjecture in affirmative for graphs with $\Delta \geq 10^{69}$.
For graphs with smaller $\Delta$, the conjecture still remains open.
It is interesting to note that the Petersen and
Hoffmann-Singleton graphs are the only two known graphs with maximum degree greater than 2, for which this bound is tight.

The second main branch of research in $L(2,1)$-labeling was pointed to analyzing the problem from the complexity point of view.
For $k \geq 4$, the $k$-$L(2,1)$-labeling problem was proven to be
NP-complete by Fiala {\it et al.} \cite{FKK} (for $k \leq 3$ the problem is polynomial). It remains
NP-complete even for regular graphs (see Fiala and Kratochv\'{i}l
\cite{fialaregular}), planar graphs (see Eggeman {\it et al.}
\cite{planar}) or series-parallel graphs (see Fiala {\it et al.} \cite{FGK}).

An exact algorithm for the so called Channel Assignment Problem, presented by Kr\'al' \cite{Kral}, implies an $O^*(4^n)$ \footnote{In the $O^*$ notation we omit polynomially bounded terms.} algorithm for the
$L(2,1)$-labeling problem. Havet {\it et al.} \cite{HKKKL} presented an algorithm for
computing $L(2,1)(G)$, which works in time $O^{*} ( 15 ^ {\frac {n}
{2} } ) = O^{*} ( 3.8730 ^ n)$. This algorithm has been improved \cite{RSiwoca,IPL},
achieving a complexity bound $O^* (3.2361^n)$.
Recently, a new algorithm for $L(2,1)$-labeling with a complexity bound $O^* (\pp ^n)$ has been presented \cite{TAMC}.

All algorithms mentioned above are based on dynamic programming approach and use exponential memory.
Havet {\it et al.} \cite{HKKKL} presented a branching algorithm for $k$-$L(2,1)$-labeling
problem with a time complexity $O^*( (k - 2.5)^n)$ and polynomial space complexity. Until now, no algorithm for $L(2,1)$-labeling
with time complexity $O(c^n)$ for some constant $c$ and polynomial space complexity has been presented.
However, there are such algorithms for a related problem of classical graph coloring. The first one,
with time complexity  $O(5.283^n)$, was shown by Bodleander and Kratsch \cite{ColoringPolySpace}.
The best currently known algorithm for graph coloring with polynomial
space complexity is by Bj\"orklund {\it et al.} \cite{incex}, using the inclusion-exclusion principle. Its time complexity
is $O(2.2461^n)$.

In this paper we present the first exact algorithm for the $L(2,1)$-labeling problem with
polynomially bounded space complexity. The algorithm works in time $O( \complexity ^n)$ (where
$\eps$ is an arbitrarily small positive constant) and is based on a divide and conquer approach.

\section{Preliminaries}

Throughout the paper we consider finite undirected graphs without
multiple edges or loops. The vertex set (edge set) of a graph $G$ is
denoted by $V(G)$ ($E(G)$, respectively).

Let $dist_G(x,y)$ be the \textit{distance} between vertices $x$ and
$y$ in a graph $G$, which is the length of a shortest path joining
$x$ and $y$.

A set $X \subseteq V(G)$ is a \textit{2-packing} in $G$ if and only
if all its vertices are in distance at least $3$ from each other
($\forall x,y \in X ~ dist_G(x,y) > 2$).

Let $N(v) = \{ u \in V(G) \colon (u,v) \in E(G) \}$ denote the set
of neighbors (the \textit{neighborhood}) of a vertex $v$. The set
$N[v] = N(v) \cup \{v\}$ denotes the \textit{closed neighborhood} of
$v$. The neighborhood of a set $X$ of vertices in $G$ is denoted by
$N(X) = \bigcup_{v \in X} N(v)$ and its closed neighborhood is
denoted by $N[X]=N(X) \cup X$.

For a subset $X \subseteq V(G)$, we denote the subgraph of $G$ induced
 by the vertices in $X$ by $G[X]$. A \textit{square} of a graph $G=(V,E)$ is the graph $G^2 = (V, \{uv \in V^2 \colon dist_G(u,v) \le 2 \})$.

\begin{definition}
For a graph $G$ and sets $Y,Z,M \subseteq V(G)$, a \emph{$(k-1)$-$L^M_Z(Y)$-labeling of a graph $G$} is a function $c \colon Y\to \{0,1,\dots,k-1\}$, such that $c^{-1}(0) \cap Z = c^{-1}(k-1) \cap M = \emptyset$, and for every $v,u\in Y$:
$$|c(v)-c(u)|\ge 2 \textrm{ if } dist_G(u,v)=1$$
$$|c(v)-c(u)|\ge 1 \textrm{ if } dist_G(u,v)=2.$$
A function $c \colon Y \to \n$ is an \emph{$L^M_Z(Y)$-labeling of $G$} if there exists $k \in \n$ such that $c$ is a $(k-1)$-$L^M_Z(Y)$-labeling of $G$
\end{definition}

\begin{definition}
For $Y,Z,M \subseteq V(G)$ let $\Lambda^M_Z(Y,G)$ denote the smallest value of $k$ admitting the existence of $(k-1)$-$L^M_Z(Y)$-labeling of $G$. We define $\Lambda^M_Z(\emptyset,G) \overset{def.}{=} 0$ for all graphs $G$ and sets $Z,M \subseteq V(G)$.
\end{definition}

Any $(k-1)$-$L^M_Z(Y)$-labeling of $G$ with $k =\Lambda^M_Z(Y,G)$ is called \textit{optimal}.

We observe that even if $c$ is an optimal $L^M_Z(Y)$-labeling of $G$, then any of the sets $c^{-1}(0)$ and $c^{-1}(\Lambda^M_Z(Y,G)-1)$ may be empty. In the extremal case, if $Z = M = Y$, then $c^{-1}(0) = c^{-1}(k-1) = \emptyset$ for all $k$ and feasible $(k-1)$-$L^M_Z(Y)$-labelings $c$ of $G$.

Notice that $\Lambda^\emptyset_\emptyset(V(G),G) = \lambda(G)+1$ for every graph $G$.

\begin{definition}
For a graph $G$, a \textit{$G$-correct partition of a set $Y\subseteq V(G)$} is a triple $(A,X,B)$, such that:
\begin{enumerate}
\item The sets $A,X,B \subseteq Y$ form a partition of $Y$
\item $X$ is a nonempty $2$-packing in $G$
\item $|A| \leq \frac{|Y|}{2}$ and $|B| \leq \frac{|Y|}{2}$
\end{enumerate}
\end{definition}

\section{Algorithm}

In this section we present a recursive algorithm for computing $\Lambda^M_Z(Y,G)$ for any graph $G$ and sets $Y,Z,M \subseteq V(G)$. It is then used to find an $L(2,1)$-span a graph $G$.

The algorithm is based on the divide and conquer approach. First, the algorithm exhaustively check if $\Lambda^M_Z(Y,G) \leq 3$. If not, the set $Y$ is partitioned into three sets $A,X,B$, which form a $G$-correct partition of $Y$. The sets $A$ and $B$ are then labeled recursively.

The labeling of the whole $Y$ is constructed from the labelings found in the recursive calls. The sets of labels used on the sets $A$ and $B$ are separated from each other by the label used for the $2$-packing $X$. This allows to solve the subproblems for $A$ and $B$ independently from each other.

Iterating over all $G$-correct partitions of $Y$, the algorithm computes the minimum $k$ admitting the existence of a $(k-1)$-$L^M_Z(Y)$-labeling of $G$, which is by definition $\Lambda^M_Z(Y,G)$.

\begin{algorithm}[H]
\caption {\algname}
\SetKwInOut{In}{Input}
\SetKwInOut{Out}{Output}
\In{Graph $G$, Sets $Y,Z,M \subseteq V(G)$}
\lIf {$Y = \emptyset$}{\Return $0$ \label{empty}}\\
\ForEach {$c \colon Y \to \{0,1,2\}$}
{\For {$k \gets 1$ \KwTo $3$}
{
    \lIf {$c$ is a $(k-1)$-$L^M_Z(Y)$-labeling of $G$}{\Return $k$} \label{smallk}
}}
$k \gets \infty$\\
\ForEach {$G$-correct partition $(A,X,B)$ of $Y$}
{
\lIf {$A \neq \emptyset$ and $B \neq \emptyset$} {$k_X \gets 1$}\label{case1}\\
\lIf {$A = \emptyset$ and $X \cap Z = \emptyset$} {$k_X \gets 1$}\label{case2}\\
\lIf {$A = \emptyset$ and $X \cap Z \neq \emptyset$} {$k_X \gets 2$}\label{case3}\\
\lIf {$B = \emptyset$ and $X \cap M = \emptyset$} {$k_X \gets 1$}\label{case4}\\
\lIf {$B = \emptyset$ and $X \cap M \neq \emptyset$} {$k_X \gets 2$\label{case5}}\\

$k_A \gets$ \textbf{\algname($G,A,Z,N(X)$)} \label{callA}\\
$k_B \gets$ \textbf{\algname($G,B,N(X),M$)} \label{callB}\\
$k \gets \min (k, k_A + k_X + k_{B} )$ \label{setk}
}
\Return $k$
\end{algorithm}

\begin{lemma}\label{lem:3color}
For a graph $G$ and sets $Y,Z,M\subseteq V(G)$, if $Y$ is a $2$-packing in $G$, then $\Lambda^M_Z(Y,G)\le 3$.
\end{lemma}

\begin{proof}
The labeling $c \colon Y \to \{0,1,2\}$ such that $c(v)=1$ for every $v\in Y$ is a $2$-$L^M_Z(Y)$ labeling of $G$. \qed
\end{proof}

\begin{theorem}
For any graph $G$ and sets $Y,Z,M \subseteq V(G)$, the algorithm call \textbf{\algname($G,Y,Z,M$)} returns $\Lambda^M_Z(Y,G)$.
\end{theorem}
\begin{proof}
If $Y = \emptyset$, the correct result is given in the line \ref{empty} (by the definition of $\Lambda^M_Z(\emptyset,G)$). If $\Lambda^M_Z(Y,G) \leq 3$, the result is found by the exhaustive search in the line \ref{smallk}. Notice that if  $|Y| \le 1$, then by Lemma \ref{lem:3color} $\Lambda^M_Z(Y,G) \leq 3$.

Assume that the statement is true for all graphs $G'$ and all sets $Y',Z',M' \subseteq V(G')$, such that $|Y'| < n$, where $n \geq 1$.

Let $G$ be a graph and $Y,Z,M$ be subsets of $V(G)$ such that $|Y|=n$. We may assume that $\Lambda^M_Z(Y,G)> 3$.
Let $k$ be the value returned by the algorithm call \textbf{\algname($G,Y,Z,M$)}.

First we prove that $k \ge \Lambda^M_Z(Y,G)$, i.e. there exists a $(k-1)$-$L^M_Z(Y)$-labeling of $G$.
Let us consider the $G$-correct partition $(A,X,B)$ of $Y$, for which the value of $k$ was set in the line \ref{setk}.
Since each of the sets  $A$ and $B$ has less than $n$ vertices, by the inductive assumption there exists a $(k_A-1)$-$L^{N(X)}_Z(A)$-labeling $c_A$ of $G$ and a $(k_B-1)$-$L^M_{N(X)}(B)$-labeling $c_B$ of $G$.

One of the following cases occurs:
\begin{enumerate}
\item If $A \neq \emptyset$ and $B \neq \emptyset$, then in the line \ref{case1} the value of $k_X$ is set to $1$ and thus $k = k_A + k_B + 1$. The labeling $c$ of $Y$, defined as follows:
\begin{displaymath}
c(v) =\left \{
\begin{array}{l l}
c_A(v)& \textrm{ if $v\in A$}\\
k_A & \textrm{ if $v\in X$}\\
k_A + 1 + c_B(v) & \textrm{ if $v\in B$}\\
\end{array}
\right.
\end{displaymath}
is  a $(k-1)$-$L^M_Z(Y)$-labeling of $G$.

\item If $A = \emptyset$ and $X \cap Z = \emptyset$, then in the line \ref{case2} the value of $k_X$ is set to $1$ and thus $k = k_B + 1$. The labeling $c$ of $Y$, defined as follows:
\begin{displaymath}
c(v) =\left \{
\begin{array}{l l}
0 & \textrm{ if $v\in X$}\\
c_B(v) + 1 & \textrm{ if $v\in B$}\\
\end{array}
\right.
\end{displaymath}
is  a $(k-1)$-$L^M_Z(Y)$-labeling of $G$.

\item If $A = \emptyset$ and $X \cap Z \neq \emptyset$, then in the line \ref{case3} the value of $k_X$ is set to $2$ and thus $k = k_B + 2$. The labeling $c$ of $Y$, defined as follows:
\begin{displaymath}
c(v) =\left \{
\begin{array}{l l}
1 & \textrm{ if $v\in X$}\\
c_B(v) + 2 & \textrm{ if $v\in B$}\\
\end{array}
\right.
\end{displaymath}
is  a $(k-1)$-$L^M_Z(Y)$-labeling of $G$.

\item If $B = \emptyset$ and $X \cap M = \emptyset$, then in the line \ref{case4} the value of $k_X$ is set to $1$ and thus $k = k_A + 1$. The labeling $c$ of $Y$, defined as follows:
\begin{displaymath}
c(v) =\left \{
\begin{array}{l l}
c_A(v) & \textrm{ if $v\in A$}\\
k_A  & \textrm{ if $v\in X$}\\
\end{array}
\right.
\end{displaymath}
is  a $(k-1)$-$L^M_Z(Y)$-labeling of $G$.

\item If $B = \emptyset$ and $X \cap M \neq \emptyset$, then in line \ref{case5} the value of $k_X$ is set to $2$ and thus $k = k_A + 2$. The labeling $c$ of $Y$, defined as follows:
\begin{displaymath}
c(v) =\left \{
\begin{array}{l l}
c_A(v) & \textrm{ if $v\in A$}\\
k_A  & \textrm{ if $v\in X$}\\
\end{array}
\right.
\end{displaymath}
is  a $(k-1)$-$L^M_Z(Y)$-labeling of $G$
 (the label $k_A+1$ is counted as used, but no vertex is labeled with it).
\end{enumerate}

The case when $X = \emptyset$ is not possible, since the partition $(A,X,B)$ is $G$-correct. The case when $A = B = \emptyset$ is not possible, since then $Y=X$ is a $2$-packing in $G$ and by the Lemma \ref{lem:3color}  $\Lambda^N_Z(Y,G)\le 3$, so the algorithm would finish in the line \ref{smallk}.

Now let us show that $k \leq \Lambda^M_Z(Y,G)$.
Let $c$ be an optimal $L^M_Z(Y)$-labeling of $G$. Let $l$ be the smallest number, such that $|c^{-1}(0) \cup c^{-1}(1) \cup \dots \cup c^{-1}(l)| \geq \frac{|Y|}{2}$.

Let $A = c^{-1}(0) \cup \dots \cup c^{-1}(l-1)$, $X = c^{-1}(l)$ and $B = c^{-1}(l+1) \cup \dots \cup  c^{-1}(\Lambda^M_Z(Y,G)-1)$.
Notice that $X$ is a $2$-packing and $X \neq \emptyset$ by the choice of $l$. Hence we observe that the partition $(A,X,B)$ is $G$-correct, so the algorithm considers it in one of the iterations of the main loop.

Let $c_A \colon A \to \n$ be a function such that $c_A(v) = c(v)$ for every $v \in A$ and $c_B \colon B \to \n$ be a function such that $c_B(v) = c(v) - (l+1)$ for every $v \in B$.
Notice that $c_A$ is an optimal $L^{N(X)}_Z(A)$-labeling of $G$ and $c_B$ is an optimal $L^M_{N(X)}(B)$-labeling of $G$, because otherwise $c$ would not be an optimal.

Hence by the inductive assumption the call in the line \ref{callA} returns the number $k_A \leq \Lambda^{N(X)}_Z(A,G)$  and the call in the line \ref{callB} returns the number $k_B \leq \Lambda^M_{N(X)}(B,G)$.

Let $k'$ be the value of  $k_A+k_X+k_B$ in the iteration of the main loop when partition $(A,X,B)$ is considered.

Let us consider the following cases:

\begin{enumerate}

\item $A,B\neq\emptyset$. In such a case the algorithm \textbf{\algname} sets $k_X=1$~in the line \ref{case1} and \\
 $\Lambda^M_Z(Y,G) = \Lambda^{N(X)}_Z(A,G) + \underbrace{1}_{c^{-1}(l)=X} + \Lambda^{N(X)}_Z(B,G) \geq k_A+k_X+k_B=k'$.

\item $A = \emptyset$ and $l=0$. In such a case  $k_A=0$ and $X \cap Z = \emptyset$ and the algorithm \textbf{\algname} sets $k_X=1$~in the line \ref{case2}  and\\
 $\Lambda^M_Z(Y,G)= \underbrace{\Lambda^{N(X)}_Z(A,G)}_{=0} +  \underbrace{1}_{c^{-1}(0)=X} + \Lambda^{N(X)}_Z(B,G) \geq  k_A+k_X+k_B=k'$.
\item  $A = \emptyset$ and $l=1$. In such a case $k_A=0$ and  $X \cap Z \neq \emptyset$. Otherwise $c'$ defined by $c'(v)=c(v)-1$ for every $v\in Y$ would be a $L^M_Z(Y)$-labeling of $G$ using less labels than the optimal $L^M_Z(Y)$-labeling $c$ of $G$ -- contradiction.
        The algorithm \textbf{\algname} sets $k_X=2$~in the line \ref{case3} and\\
 $\Lambda^M_Z(Y,G) = \underbrace{\Lambda^{N(X)}_Z(A,G)}_{=0} + \underbrace{1}_{c^{-1}(0)=\emptyset} +  \underbrace{1}_{c^{-1}(1)=X} + \Lambda^{N(X)}_Z(B,G) \geq  k_A+k_X+k_B=k'$.

\item $B = \emptyset$ and $l=\Lambda^M_Z(Y,G)-1$. In such a case $k_B=0$ and  $X \cap M = \emptyset$, and the algorithm \textbf{\algname} sets $k_X=1$~in the line \ref{case4}  and\\
$\Lambda^M_Z(Y,G) = \Lambda^{N(X)}_Z(A,G) +  \underbrace{1}_{c^{-1}(\Lambda^N_M(Y,G)-1)=X} + \underbrace{\Lambda^{N(X)}_Z(B,G)}_{=0} \geq k_A+k_X+k_B=k'$.
\item  $B = \emptyset$ and $l=\Lambda^M_Z(Y,G)-2$. In such a case $k_B=0$ and  $X \cap M \neq \emptyset$ and the algorithm \textbf{\algname} sets $k_X=2$~in the line \ref{case5} and \\
 $\Lambda^M_Z(Y,G) = \Lambda^{N(X)}_Z(A,G) +  \underbrace{1}_{c^{-1}(\Lambda^N_M(Y,G)-2)=X} + \underbrace{1}_{c^{-1}(\Lambda^N_M(Y,G))=\emptyset} + \underbrace{\Lambda^{N(X)}_Z(B,G)}_{=0} \geq k_A+k_X+k_B= k'$.
\end{enumerate}
Since those are all possible cases and $k$ is the minimum over values of $k'$ for all correct partitions, clearly $k \leq \Lambda^M_Z(Y,G)$.
\qed
\end{proof}

\begin{observation}\label{obs:lambda}
By the definition of $\Lambda^M_Z(Y,G)$, the algorithm call \textbf{\algname}($G,V(G),\emptyset,\emptyset$) returns $\lambda(G)+1$.
\end{observation}

\begin{lemma}\label{lem:compl}
Let $G$ be a graph on $n$ vertices, $Y,Z,M \subseteq V(G)$ and let $y = |Y|$. If $G^2$ is computed in advance, the algorithm \textbf{\algname} finds  $\Lambda^M_Z(Y,G)$ in the time $O(C^{\log y}y^{3\log y}9^y)$ and polynomial space, where $C$ is a positive constant.
\end{lemma}

\begin{proof}
Having $G^2$ computed, checking if any two vertices in $V(G)$ are in distance at most $2$ from each other in $G$ takes a constant time. Hence verifying if a given $X \subseteq Y$ is a $2$-packing in $G$ can be performed in the time $O(y^2)$. Moreover, we can check if a given function $c \colon Y \to \n$ is an $L^M_Z(Y)$-labeling of $G$ in the time $O(y^2)$.

Let $y = |Y|$ be the measure of the size of the problem.
Let $T(y)$ denote the running time of the algorithm \textbf{\algname} applied to a graph $G$ and $Y,Z,M\subseteq V(G)$.

Ale algorithm \textbf{\algname} first checks in constant time if $Y = \emptyset$. Then it exhaustively checks if there exists a $(k-1)$-$L^M_Z(Y)$-labeling of $G$ for $k \in \{1,2,3\}$. There are $3^y$ functions $c \colon Y \to \{0,1,2\}$, so this step is performed in the time $O(y^2 \cdot 3^y)$.

Then for every $G$-correct partition of $Y$ the algorithm is called recursively for two sets of size at most $\frac{y}{2}$. Notice that there are at most $3^y$ considered partitions. Checking if a partition of $Y$ is $G$-correct can be performed in time $O(y^2)$. Hence we obtain the following inequality for the complexity ($C_1$ and $C_2$ are positive constants):
$$T(y)\leq C_1y^24^y+C_2y^33^y2\cdot T \left (\frac{y}{2} \right ) $$
Let $C=\max(C_1,2C_2)$, then
$$T(y)\leq Cy^24^y+Cy^33^y\cdot T \left (\frac{y}{2} \right ) $$
It is not difficult to verify that $T(y)\leq \complexityy = O(C^{\log y}y^{3\log y}9^y)$, where $D$ is a positive constant.

The space complexity of the algorithm is clearly polynomial. \qed
\end{proof}

\begin{theorem}
For a graph $G$ on $n$ vertices $\lambda(G)$ can be found in the time  $O(\complexity^n) $ and polynomial space, where $\eps$ is an arbitrarily small positive constant.
\end{theorem}

\begin{proof}
The square of a graph $G$ can be found in the time $O(n^3)$.
By the Observation \ref{obs:lambda} and Lemma \ref{lem:compl}, the algorithm \textbf{\algname} applied to $G$, $Y=V(G)$ and $Z=M=\emptyset$ finds $\Lambda^\emptyset_\emptyset(V(G),G)=\lambda(G)-1$ in the time $O(C^{\log n}n^{3\log n}9^n)=O(\complexity^n)$ and polynomial space. \qed
\end{proof}

\subsubsection*{Remark}
We have just learned that results similar to those included in this paper were independently obtained (but not published) by Havet, Klazar, Kratochv\'{i}l, Kratsch and Liedloff \cite{HKKKLpolyspace}.

\end{document}